\documentstyle[epsfig,aps,floats,twocolumn,prb]{revtex}

\begin{document}

\author{Thorsten Dr\"ose and Cristiane Morais-Smith}
\title{Crossovers in the thermal decay of metastable states
       in discrete systems} 
\address{ I. Institut f\"ur Theoretische Physik, Universit\"at Hamburg,
Jungiusstrasse~9, D-20355 Hamburg, Germany \\
and \\
Institut de Physique Th\'eorique, 
Universit\'e de Fribourg, P\'erolles, CH-1700 Fribourg, Switzerland \\
\rm{(\today)}\bigskip\\
\parbox{14.4cm}{\rm
The thermal decay of linear chains from a metastable state is 
investigated. A crossover from  rigid to elastic decay
occurs when the number of particles, the single-particle energy barrier, or
the coupling strength between the particles is varied. 
In the rigid regime, the single-particle energy barrier is small compared
to the coupling strength, and  the decay occurs via a uniform 
saddle-point solution, with all degrees of freedom decaying instantly.
Increasing the barrier one enters the elastic regime, where
the decay is due to bent saddle-point 
configurations using the elasticity of the chain to lower their 
activation energy. Close to the rigid-to-elastic crossover, 
nucleation occurs at the boundaries of the system.
However, in large systems, a second crossover
from boundary to bulk nucleation can be found
within the elastic regime, when the single-particle energy barrier is further 
increased. 
We compute the decay rate in the rigid and  elastic regimes
within the Gaussian approximation. Around the rigid-to-elastic crossover,
the calculations are performed beyond the steepest-descent approximation.
In this region, the prefactor exhibits a scaling property.
The theoretical results are discussed in the context of discrete Josephson
transmission lines and pancake vortex stacks  that are pinned by columnar 
defects. }}

\maketitle
\section{Introduction}
The decay of metastable states in systems with one \cite{Kram40} or more
degrees of freedom \cite{Lang67-69}(DOF) 
has been intensively studied in the last decades \cite{Haen90}.
The crossover from rigid to elastic decay \cite{Ivle87,Lefr92,Mora94} 
was studied in systems with one and two DOF by using methods 
known from the  analysis of the crossover from thermal to quantum 
decay \cite{Affl81,Lark83,Grab84a,Weis99}. 
In this work we consider a system with $N$ 
DOF and investigate crossovers that occur in its thermal decay
from a metastable state while tuning an external parameter. 

A system localized in a relative 
minimum of a potential energy surface can escape
from the trap due to thermal or quantum fluctuations. At high temperatures
the decay process is purely thermal, and most probably 
occurs through the free-energy lowest-lying saddle point
that connects two local minima. In this paper we study a model where 
the energy surface changes upon varying an external parameter
$ \delta $. Above a critical value $ \delta_{*}$, the saddle point bifurcates 
into new lower lying ones, 
causing an enhancement of the escape rate $ \Gamma $.
In the steepest-descent approximation $ \Gamma(\delta)$  and its derivative 
$ \Gamma'(\delta)$ are  continuous at $ \delta_{*} $, 
whereas the second derivative $ \Gamma''(\delta_{*})$ diverges. 
This behavior can be interpreted in terms of a second-order phase 
transition \cite{Lark83} and hence is called crossover of second order.

Experimental measurements concerning the decay of metastable states 
in dc superdonducting interference devices (SQUID's) were interpreted in terms of a saddle-point splitting of the 
potential energy \cite{Lefr92}. This device consists
of a superconducting ring intercepted by two Josephson junctions (JJ's).
The phase differences across the junctions play the role of generalized 
coordinates. 
The inductance of the circuit couples the two phases. By reducing the bias
current $I$ that flows through the system, the decay of the phases changes
from a rigid regime \cite{Yong86,Shar88,Han89}, 
with the two phases decaying together as if they were rigidly coupled, 
to an elastic  regime,
with the phases decaying independently \cite{Lefr92}.

An interesting question is how such a crossover occurs in more complex
systems like in a discrete Josephson transmission line (DJTL), which is
a one-dimensional array of $N$ parallelly coupled JJ's. Instead of two DOF,
one would then have $N$ coupled DOF. Another example of such
a system is a stack of $N$ pancake vortices \cite{Clem91} in a 
layered superconductor in the presence of columnar defects
\cite{Blat94}. A vortex pinned by a columnar defect,
but subject to a driving current flowing perpendicular to the magnetic field 
can escape from the trap by thermal activation. 
The open question is then whether a transition from a
rigid  to an elastic  behavior can
be found in the vortex or the DJTL systems, and also if more crossovers
inside the elastic regime would arise due to the different decay 
possibilities involving the large number of DOF. 
In this paper we analyze the  crossover in the decay process
due to a saddle-point bifurcation in systems with $ N>2$
DOF. It turns out that for $ N=3 $ the saddle points of the
potential energy can still be solved exactly. For larger $ N $ we
determine them perturbatively. Furthermore, we find that for $ N \gg 1$
a second crossover from  boundary to bulk nucleation
can take place in the elastic regime.

The thermal escape rate $ \Gamma_{th} = P \exp(-U_{a}/k_{B}T)$ 
is determined in the rigid and 
elastic regimes for an arbitrary number of particles, 
by assuming an overdamped motion  out of  a weakly metastable state.  
Far from the saddle-point bifurcation, $ \Gamma_{th} $ is
evaluated within the Gaussian approximation, including the pre-exponential 
factor $P$. Close to the crossover from rigid decay to boundary 
nucleation, we calculate the rate beyond steepest descent
and find that $ P $ displays a scaling property.

The paper is organized as follows: In Sec.\ II we introduce the model
that can be applied both to DJTL and to pancake vortices in layered 
superconductors in the 
presence of a columnar defect. In Sec.\ III  we determine 
the crossover from rigid decay to elastic boundary nucleation
and the corresponding decay diagram. We show that in the elastic regime a 
second crossover from boundary to bulk nucleation can occur.
We evaluate the saddle-point 
solutions and their activation energies in the three decay regimes. 
In Sec.\ IV the thermal escape rate is calculated.
Finally, we discuss our results and draw our conclusions in Sec.\ V.

\section{Model}
\subsection{Free energy}
Let us consider a system of $ N $ degrees of freedom 
$ {\bf u} = (u_0,\dots, u_{N-1})$, each of them experiencing
a single-particle potential $ U(u_{n}) $, and interacting with one 
another via spring-like nearest-neighbor interactions,
\begin{equation}\label{eff_free}
{\cal E}( {\bf u}) = 
\frac{\kappa}{2} \sum_{n=1}^{N-1} ( u_{n}-u_{n-1})^{2} 
+ \sum_{n=0}^{N-1} U(u_{n}),
\end{equation}
where $ \kappa $ is the spring constant.
We assume that all the 
particles are initially situated near a local minimum of the potential
$ U $. The  coordinates $ u_{n} $ measure the distance of each particle 
$n$ from this minimum.
Close to the local minimum $ u_{n} = 0$, the single-particle potential can be
approximated by a cubic parabola,
\begin{equation} \label{single_pot}
U(u_{n}) = U_{B} \left[ 3 \delta \left( \frac{ u_{n} }{R} \right)^{2} 
  - 2  \left( \frac{ u_{n} }{R} \right)^{3}  \right]. 
\end{equation}
Here $ \delta \ll 1$ is a tunable parameter. The constants $ U_{B} $ and
$ R $ are the characteristic energy and length scales, respectively.
At $ u_{n}= R \delta $ the single-particle potential has a maximum. 
The energy difference between the local minimum and the maximum is 
$ {\tilde U_{B}}= U(R \delta) = U_{B} \delta^{3}$.

\begin{figure}[t]
\unitlength1cm
\begin{picture}(8,3.5)
\epsfxsize=7.5cm
\put(0.5,0.5){\epsfbox{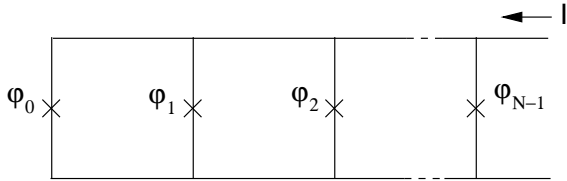}}
\end{picture}
\caption[]{\label{DJTL} 
A current biased one-dimensional array of identical 
parallelly coupled Josephson junctions,
also called  a discrete Josephson transmission line. The relevant
degrees of freedom are the phase differences $ \varphi_{n}$ across
the junctions.}
\end{figure}
Among the physical systems that can be described by 
Eq.\ (\ref{eff_free}) are the DJTL,
see Fig.\ \ref{DJTL}.
The potential energy of a system of $ N $ identical JJ's in the 
presence of a bias current $ I $ is
\begin{eqnarray} \label{static}
V(\varphi_0,\dots,\varphi_{N-1})& =& 
\frac{E_{J}^{2}}{2L I_{c}^{2} } 
\sum_{n=1}^{N-1} (\varphi_{n} -\varphi_{n-1})^{2} 
\nonumber \\
& & + E_{J} \sum_{n=0}^{N-1} \left[1-\cos(\varphi_{n})
-\frac{I   \varphi_{n}}{N I_{c}} \right]. 
\end{eqnarray}
Here the phase differences across the JJ's are given by 
$  \varphi_0,\dots,\varphi_{N-1}$. 
The first term in Eq.\ (\ref{static}) represents the interaction energy due 
to the inductances 
between the loops. Here only the self-inductances of the loops are 
taken into account, whereas the mutual inductances are neglected \cite{Bock94}.
The elastic constant is $ \kappa = E_{J}^{2}/ L I_{c}^{2} $,
where  $ E_{J} = (\Phi_{0}/2 \pi) I_{c}$ is the Josephson energy, 
$ I_{c} $ the critical current of a 
single junction, $ L $ is the inductance, and 
$  \Phi_{0} = hc/2e$ is the flux quantum.
The second term represents the tilted washboard potentials of the driven JJ's  
that arise due to the relation between currents and gauge invariant phases  
across the junctions.  If we concentrate on the experimentally most 
interesting limit of currents $I$ close to criticality, 
$ N I_{c} - I \ll N I_{c} $,
the tilted washboard potential can be well approximated by its 
cubic expansion, and we can identify 
$ {\cal E} = V $ 
with $ u_n =  \varphi_n + R \delta/2 - \pi/2 $, 
 $ U_{B} = 4 \sqrt{2} E_{J}/3 $, 
$ R = 2 \sqrt{2},$ and 
$\delta = \sqrt{(1-I/NI_{c})}$.

\begin{figure}[t]
\unitlength1cm
\begin{picture}(8,4.5)
\epsfxsize=5cm
\put(1.5,0.5){\epsfbox{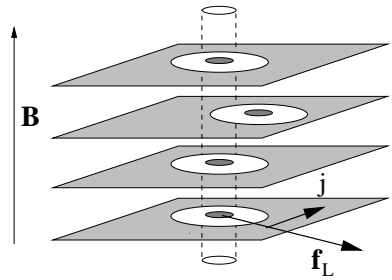}}
\end{picture}
\caption[]{\label{model}
A stack of ``pancakes'' produced by a magnetic field $ {\bf B} $ applied
perpendicular to the layers. The pancakes are coupled to each other via
magnetic interaction and Josephson currents.
A columnar defect pins the vortex. When a current
$ j $ is flowing through the system, a Lorentz force  $ {\bf f}_{L}$ acts on
the pancakes, reducing the energy barrier the vortex has to overcome
to escape from the defect.}
\end{figure}

Another physical realization of the model described by Eq.\ (\ref{eff_free})
is a stack of pancake vortices trapped in a columnar defect, 
which is artificially introduced in a layered superconductor.
Both the magnetic field
that produces the pancake vortices and the columnar defect are
perpendicular to the superconducting layers; see Fig.\ \ref{model}.
Once a bias current $  {\bf j} = j {\bf e}_{y} $ flows through the layers 
perpendicular to the magnetic field pointing
in the $z$ direction, the pancakes will be driven by the resulting 
Lorentz force. The corresponding free energy reads 
\begin{equation}\label{Veff}
{\cal F}  = 
 \frac{\varepsilon_{l}}{2s} 
\sum_{n=1}^{N-1} ({\bf u}_{n}-{\bf u}_{n-1})^{2} 
+
\sum_{n=0}^{N-1} \left[ U_{p}({\bf u}_{n}) - 
{\bf f}_{L} \cdot {\bf u}_{n}) \right].
\end{equation}
The
displacement of the $n$th pancake vortex from its equilibrium 
position in the columnar defect is now given by a two-dimensional
vector $ {\bf u}_{n}=(u_{n,x}, u_{n,y})$.
The first sum in Eq.\ (\ref{Veff}) models the magnetic 
and Josephson couplings 
between the layers by  elastic interactions between pancakes in
adjacent layers \cite{Kosh96}. Here 
$ \varepsilon_{l} = 
(\varepsilon_{0}/\gamma^{2}) \ln(\lambda_{ab} / \xi_{ab})$ 
is the elastic constant, 
$ \varepsilon_{0} = \Phi_{0}^{2}/(4\pi \lambda_{ab})^{2}$ is the vortex self 
energy, $ \gamma = \lambda_{c} / \lambda_{ab} $ is the anisotropy ratio of 
the penetration depths $ \lambda_{c} $ and $ \lambda_{ab}$, $ s$ is the 
interlayer spacing, and $ \xi_{ab} $ is the in-plane coherence length. 
The second sum contains the columnar defect pinning potentials  $ U_{p} $
felt by the single pancakes and 
the Lorentz force density
$  {\bf f}_{L} = \Phi_{0}~ {\bf j} \wedge {\bf e}_{z} / c$,
where $ {\bf e}_{z} $ is the unit vector pointing perpendicular to the 
planes. The potential $ U_{p} $ 
is smooth on the length scale $ \xi_{ab}$ with a local minimum at the 
center of the defect. An upper estimate for the depth of the potential well 
is given by $ U_{B} \approx t \varepsilon_{0} \ln(R/\xi_{ab}) $, 
where $ R $ is the radius of the columnar defect \cite{Blat94}. 
The parameter $t$ denotes the superconducting layer thickness.
In  the large current limit,
$ \delta = \sqrt{1-j/j_{c}} \ll 1$ gives a measure of how close the 
current $ j $ is to the critical current $ j_{c}$.  
Then the sum of the  pinning and the Lorentz part of the free energy
is approximately
\begin{equation} \label{approx_pot}
U_{B} \sum_{n=0}^{N-1} \left[ 3 \delta \left( \frac{ u_{n,x} }{R} \right)^{2} 
 - 2  \left( \frac{ u_{n,x} }{R} \right)^{3} 
 + \frac{3}{2} \left( \frac{ u_{n,y} }{R} \right)^{2}
\right],
\end{equation}
where we have kept 
only the terms that are of order $ \delta^{3}$. 
The terms proportional to 
$ \delta (u_{n,y}/R)^{2} $ and  $ u_{n,y}^{2} u_{n,x} /R^{3}$,
that are of the order $ \delta^{4}$, have been neglected.
Hence the displacements in the $y$ direction are 
essentially decoupled from the displacements in the $x$ direction.
As a consequence, two identical integrals over $u_{n,y}$ appear in the 
enumerator and in the denominator of the decay rate expression, \cite{Lang67-69}
which will cancel each other.
For this reason, we will neglect $u_{n,y}$ in the following.
Renaming  $ u_{n}= u_{n,x}$, we obtain Eq.\ (\ref{eff_free})
with $ \kappa = \varepsilon_l/s$.

\subsection{Decay rate} \label{rate}
Well above the crossover temperature $ T_{0}$ that separates the 
thermally activated decay regime from the quantum tunneling regime, 
$ T \gg T_{0}$, the escape of the DOF from the pinning potential can be 
described by a Langevin equation 
$ \eta \stackrel{.}{\bf u} + \nabla {\cal E}({\bf u}) = {\bf f}(t),$
assuming that the motion is overdamped. 
Here $ \eta $ denotes the friction coefficient. 
If we consider the resistively shunted model for the DJTL, 
$ \eta $ is the inverse shunting resistance.
For the vortex problem, $ \eta $ is given by the
Bardeen-Stephen coefficient \cite{Bard65}.              

The white noise random force $ {\bf f}(t)$ represents a heat bath at 
temperature $T$. It has ensemble averages $ \langle f_{i}(t) \rangle = 0 $ 
and $ \langle f_{i}(t) f_{j}(t') \rangle=
2\eta k_{B} T \delta_{ij} \delta(t-t') $.
In the limit of weak metastability, where the barrier is much
larger than the thermal energy $  U_{a} \gg k_{B}T $,
the corresponding Klein-Kramers equation can be reduced to a 
Smoluchowsky equation \cite{Kram40}.
The escape rate $ \Gamma_{th}$ for the (quasi)stationary 
case was determined to be  \cite{Lang67-69}
\begin{equation} \label{thermal_rate1}
\Gamma_{th} = \frac{1}{\eta} \left( \frac{k_{B} T |\mu_{0}^{s}|}{2 \pi}
\right)^{1/2}
\frac{\int_{\cal S} d^{N-1} {\bf u'}  
~{\rm e}^{ -{\cal E} ({\bf u'})  /k_{B}T  } }
{\int_{\cal V} d^{N} {\bf u} 
~{\rm e}^{ -{\cal E}({\bf u} )/k_{B}T  } },
\end{equation}
where $ {\bf u'} \in {\cal S}$, $ {\bf u} \in {\cal V}$,
$ {\cal S}$ is the hypersurface in the configuration space 
intersecting  the saddle point(s) perpendicular to the unstable 
direction(s), $ {\cal V}$ is the configuration volume occupied by 
all metastable solutions and $ \mu_{0}^{s} $ is the curvature of the energy
surface ${\cal E}({\bf u} ) $ along the unstable direction evaluated at the
saddle point.

Solving Eq.\ (\ref{thermal_rate1}) in the steepest-descent 
approximation, one can derive the Arrhenius law
\begin{equation} \label{thermal_rate}
\Gamma_{th}= P(\delta)~ \exp \left( -\frac{U_{a}(\delta)}{k_{B}T} \right).
\end{equation} 
The activation energy $ U_{a} $ is obtained by evaluating the energy
functional (\ref{eff_free}) at the saddle-point configuration, which 
will be done in Sec. \ref{sec_sad}. 
The computation of the prefactor $P$
is a more involving task. In this case, we have to analyze the spectrum
of the curvature matrix $ \partial_{n}  \partial_{m} {\cal E} $
at the minimum and at the saddle point, since $P$ describes the contributions 
to the rate that stem from the fluctuations around the extrema.   
At a characteristic value $ \delta = \delta_{*}$ the saddle 
bifurcates indicating a crossover from a rigid regime to an elastic regime.
In the crossover region, 
the steepest-descent approximation cannot be applied. However, even beyond 
the steepest-descent approximation, the form of Eq.\ (\ref{thermal_rate})
remains valid.  
The calculations of $P$ will be performed in Sec. \ref{sec_pre}.

\section{Saddle-point solutions and their activation energies} 
\label{sec_sad}
The thermally activated
escape from the local minimum $  {\bf u}_{min}=(0,\dots,0) $
of the potential proceeds mainly via the saddle-point
solutions ${\bf u}_{s} $ of (\ref{eff_free}).
These  unstable stationary solutions satisfy
$ \nabla_{u} {\cal E}( {\bf u}_{s})=0$,
and their curvature matrix $ {\bf H}( {\bf u}_{s} )$ with 
elements 
\begin{equation} \label{hesse}
H_{nm}({\bf u}_{s}) = 
\frac{\partial^{2}}{\partial u_{n} \partial u_{m}} {\cal E}({\bf u}_{s})
\end{equation}
has at least one negative eigenvalue.

\subsection{Saddle-point bifurcation} \label{bifurcation}
The saddle point $ {\bf u}_{rs} = (R \delta,\dots,R \delta)$,
which we call the rigid saddle point (rs),
can be readily identified. 
In Appendix A we calculate the eigenvalues of a curvature matrix
for a uniform extremal solution. 
Using Eq.\ (\ref{eigenss}) we find the eigenvalues for $ H({\bf u}_{rs}) $,  
 \begin{equation} \label{eigen_sad}
\mu_{n}^{rs} =  - \frac{6 U_{B} \delta}{R^{2}}
     + 4 \kappa\sin^{2} \left( \frac{n \pi}{2 N} \right). 
\end{equation}
The lowest eigenvalue $ \mu_{0}^{rs} =- 6 U_{B} \delta /R^{2}  < 0$ 
indicates that there is at least 
one unstable direction. It is the only one, if $ \delta $ is smaller than
\begin{equation}
\delta_{*} = \frac{2 \kappa R^{2}}{3 U_{B}} 
\sin^{2} \left( \frac{\pi}{2N} \right).
\end{equation}
\begin{figure}[t]
\unitlength1cm
\begin{picture}(8,7.)
\epsfxsize=7cm
\put(0.5,0.5){\epsfbox{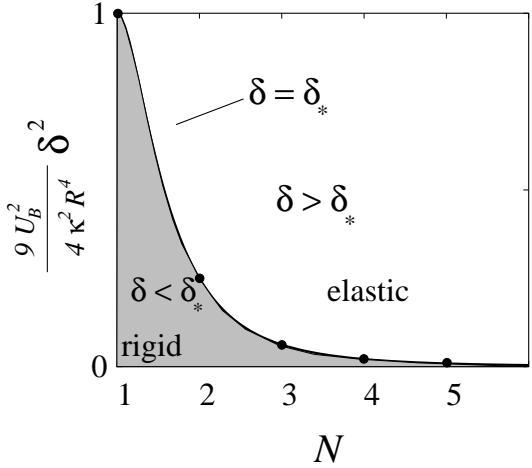}}
\end{picture}
\caption[]{\label{creep_dia} The decay diagram of a system
with a small number of degrees of freedom $ N$. 
The solid line and the dots indicate the 
crossover from rigid to elastic decay at $ \delta = \delta_{*}$ 
as a function of $N$. }
\end{figure}
However, when  $ \delta \to \delta_{*}$, 
the eigenvalue $ \mu_{1}^{rs} = 6 U_{B}
(\delta_{*}-\delta) /R^{2}$ vanishes.
At $ \delta = \delta_{*} $ the saddle splits 
indicating  the  existence of an {\em elastic} saddle-point configuration 
$ {\bf u}_{es}$.
Below, we will show that for $ \delta > \delta_{*} $ 
the  energy ${ \cal E} ({\bf u}_{es})$ 
is smaller than $ { \cal E} ({\bf u}_{rs})= N {\tilde U}_{B} $.
Hence the {\em elastic} saddle-point 
configuration  $ {\bf u}_{es}$ instead of the {\em rigid} one is the most 
probable configuration that leads to decay. One identifies 
the energy of the most probable 
configuration with the activation energy $ U_{a}$.
The saddle-point bifurcation can thus be interpreted as a 
crossover between two types of decay:
the crossover from a {\em rigid} regime with an activation
energy $ U_{a} (\delta \le \delta_{*} ) = N {\tilde U}_{B} $ 
to an {\em elastic} regime with
$ U_{a} (\delta > \delta_{*} ) = { \cal E} ({\bf u}_{es})$. 
The corresponding decay diagram is shown 
in Fig.\ \ref{creep_dia}.

\subsection{Rigid and elastic saddles} \label{saddles}
We now calculate the elastic saddle-point solutions. First, we discuss
the appearance of the elastic saddle in the crossover regime
for arbitrarily many DOF. The evolution of the elastic saddle point 
with increasing $ \delta $ is elucidated by analyzing the exactly solvable
case of three DOF. Far from the crossover, the three-particle result is used 
to make an ansatz for the $N$-particle solutions, which can again be 
determined perturbatively.

Near the crossover, we expand the elastic solution 
around the rigid one, $ {\bf u}_{es} =  {\bf u}_{rs} +\Delta {\bf u} $. 
Then $ {\cal E} $ is most conveniently represented in the coordinate system 
of the principal axis of $ H({\bf u}_{rs})$, where  $ {\cal E}$ is 
diagonal in the coordinates up to second order. 
The transformation is 
achieved by rewriting $ \Delta {\bf u} $ as a trigonometric polynomial,
\begin{equation} \label{trans}
u_{n} = R \delta \left[ 1 + \sum_{k=0}^{N-1} q_{k} 
\cos \left( \frac{\pi k \left( n+ 1/2 \right)}{N} \right)  \right]. 
\end{equation}
Here the coordinates 
$ q_{k} $ are the dimensionless amplitudes of the Fourier modes with a wave 
number $k$ that measure the deviations from the rigid saddle-point 
solution $ u_{n}^{rs} = R \delta $. 
In this coordinate system, the energy functional reads
\begin{eqnarray}
{\cal E}( {\bf q})
& = & N {\tilde U}_{B} \left[ \
     1- 2 q_{0}^{3}    
  + \frac{1}{2}\sum_{k=0}^{N-1} {\tilde \mu_{k}}^{rs} q_{k}^{2} \right.  \\
&&       - 3 q_{0} \sum_{k=1}^{N-1} q_{k}^{2} 
    -\frac{1}{2} \sum_{k=1}^{N-1}q_{k}^{2}  \left( q_{2k} - q_{2(N-k)} \right)
\nonumber \\
&& \left.   - \sum_{m>k=1}^{N-1} 
     q_{m} q_{k} \left( q_{m+k} +q_{m-k} - q_{2N-m-k}\right) 
   \right] , \nonumber
\end{eqnarray}
where we define $ q_{k} \equiv 0 $ for $ k<0 $ or $ k \ge N$,
and  $ {\bf q}=(q_{0},\dots, q_{N-1})$. In  the new
coordinate system, the dimensionless eigenvalues $ {\tilde \mu_{k}} $ of the
curvature matrix are given by 
$ {\tilde \mu_{0}} = (R^{2}/U_{B} \delta) \mu_{0}$ and
$ {\tilde \mu_{k}} = (R^{2}/2 U_{B}\delta) \mu_{k}$ for $ k \not= 0$. 
The different
prefactors are due to transformation (\ref{trans}).
At the rigid saddle one finds
\begin{equation}
{\tilde \mu_{k}}^{rs} =
 \frac{2 \kappa R^{2}}{U_{B}  \delta}  
   \sin^{2}\left( \frac{\pi k}{2N}\right)  - 3 - 3\delta_{0,k}, 
\end{equation}
where $\delta_{0,k}$ is the Kronecker delta function.
For $ \delta < \delta_{*} $, 
where the saddle-point solution is the rigid one with 
$ u_{n} = R \delta$, all the values $ q_{n} =0$. 
The second order expansion of $ {\cal E}$ around 
the rigid saddle point reads
\begin{equation}
{\cal E}( {\bf q}) =  N {\tilde U}_{B}
\left( 1+ \frac{1}{2} \sum_{k=0}^{N-1} {\tilde \mu_{k}}^{rs} 
q_{k}^{2} \right).
\end{equation}
At the crossover, $ {\tilde \mu_{1}}^{rs}$ vanishes 
and the quadratic approximation
of $ {\cal E} $ becomes independent of $ q_{1}$. Since large
fluctuations in $ q_{1}$ would not contribute to the free energy, this 
approximation becomes insufficient within the crossover regime 
where $ \ {\tilde \mu_{1}}^{rs} \ll 1 $. 
Thus, in order to describe the free energy contributions of fluctuations in $
q_{1} $ more properly, higher-order terms in $ q_{1}$ that arise due to the 
coupling to the other fluctuation coordinates have to be taken into account.
One estimates that 
$ \Delta^{2} {\cal E} \sim q_{n \not = 1}  q_{m \not = 1} 
\sim  q_{n \not = 1} q_{1}^{2} $. In comparison, the third-order terms 
$ q_{k}  q_{m}  q_{n}$ with $ m,n \not= 1$ are much smaller and
hence can be  neglected. Since $ q_{1}^{2} $ is only coupled to 
$ q_{0}$ and $ q_{2}$, one finds 
\begin{equation} \label{beyond1}
{\cal E}( {\bf q}) = N {\tilde U}_{B} 
\left[1+ \frac{1}{2} \sum_{k=0}^{N-1}{\tilde \mu}_{k}^{rs} q_{k}^{2}  
       - 3 q_{1}^{2} \left(q_{0} + \frac{q_{2}}{2}  \right)   \right]. 
\end{equation}
In the following, we define the small parameter  
$ \epsilon = (1- \delta_{*}/\delta)= -{\tilde \mu_{1}}^{rs}/3$, 
which measures the distance from the crossover. It is positive in the 
elastic regime and negative 
in the rigid one. Within the crossover regime, $ -1 \ll \epsilon \ll 1 $. 
By solving $ \nabla {\cal E} = 0$, one finds the extrema.
In addition to the extrema already found  in the rigid regime, 
an  elastic saddle-point 
solution  $ {\bf q}_{es}$ with a {\em single} kink emerges slightly below 
the crossover, for $ \delta > \delta_{*}$,
\begin{eqnarray} \label{approx_ela}
q_{0}^{es} & = & \frac{9 \epsilon }{2D {\tilde \mu_{0}}^{rs}}, \nonumber \\
q_{1}^{es} & = & \left( \frac{3 \epsilon}{2D} \right)^{1/2}, \\
q_{2}^{es} & = & \frac{9 \epsilon}{4D {\tilde \mu_{2}}^{rs}},  \nonumber \\
q_{k}^{es} & = & 0, ~k>2  \nonumber,
\end{eqnarray}
where ${\tilde \mu_{0}}^{rs},~ {\tilde \mu_{2}}^{rs}$, and 
\begin{equation} \label{D}
D = - 18/(2 {\tilde \mu_{0}^{rs}}) - 9/(4 {\tilde \mu_{2}^{rs}}) 
= 3/2 - 9/(4 {\tilde \mu_{2}^{rs}}) 
\end{equation}
are evaluated at the crossover.
This {\em elastic} solution has a lower activation energy 
$ U_{a}^{es} \approx 
N U_{B} \delta^{3}  [1- C \epsilon^{2}(\delta)]$  
than the stiff solution. Here $ C=(54-81/{\tilde \mu_{2}}^{rs})/32D^{2} $ 
is a positive constant of the order of unity, 
since ${\tilde \mu_{2} }^{rs}\ge 6 $ for $ N \ge 3$.
Since both $ U_{a}(\delta ) $ and its derivative 
$  U_{a}'(\delta) $ are continuous, 
but $  U_{a}''(\delta) $ is discontinuous at $ \delta = \delta_{*}$,
the crossover from rigid to elastic decay is of second order. 

In order to illustrate that in our discrete model, 
close to the crossover, {\em boundary} nucleation
is the dominant process leading to decay in the elastic regime, 
we will study a chain  consisting of three particles, where the saddle-point 
solutions can be determined exactly. 
The parameter $\epsilon $ can now take any value in the interval 
$ -\infty < \epsilon \ll 1- \delta_{*}$.
After substituting $ {\tilde \mu_{0}}^{rs} = -6, 
{\tilde \mu_{1}}^{rs} = -3 \epsilon $ and 
$ {\tilde \mu_{2}}^{rs} = 6 - 9 \epsilon $, the free-energy function reads
\begin{eqnarray}
{\cal E}(q_{0},q_{1},q_{2})
& = & 3 {\tilde U}_{B} \left[ \left.\frac{}{}
1- 3 q_{0}^{2} - 2 q_{0}^{3} \right. \right.  \nonumber \\ 
&& - \frac{3 \epsilon}{2}  q_{1}^{2}
 + \left(3 - \frac{9 \epsilon}{2} \right) q_{2}^{2}  \\
&& \left. -3 q_{0} \left(q_{1}^{2} + q_{2}^{2} \right)
- \frac{q_{2}}{2} \left( 3 q_{1}^{2} - q_{2}^{2} \right)
\right], \nonumber
\end{eqnarray}
From the extremal condition $ \nabla {\cal E} = 0 $ we 
calculate the extrema and find that
slightly below the crossover in the elastic regime, only 
${\bf q}_{es} = (q_{0}^{es},q_{1}^{es},q_{2}^{es})$, with
\begin{eqnarray}
q_{0}^{es}& = &-\frac{2 \epsilon }{3}, \\
q_{1}^{es} &=& (\pm) \left( \frac{4 \epsilon }{3} - \epsilon^{2} 
\right)^{1/2}, \\
q_{2}^{es} & = & \frac{\epsilon}{3}, 
\end{eqnarray}
is a possible elastic saddle-point solution. 
Energetically, the sign in front of $q_{1}^{es} $ does not have any 
relevance since $ q_{1} $ appears only quadratically in 
$ {\cal E}$. It arises due to the existence of two 
degenerate solutions that can mapped into each other by changing the sign of 
$ q_{1}$, which is equivalent to a mirror symmetry transformation. 
\begin{figure}[t]
\unitlength1cm
\vspace{1cm}
\begin{picture}(8,6.2)
\epsfxsize=7cm
 \put(0.5,0.5){\epsfbox{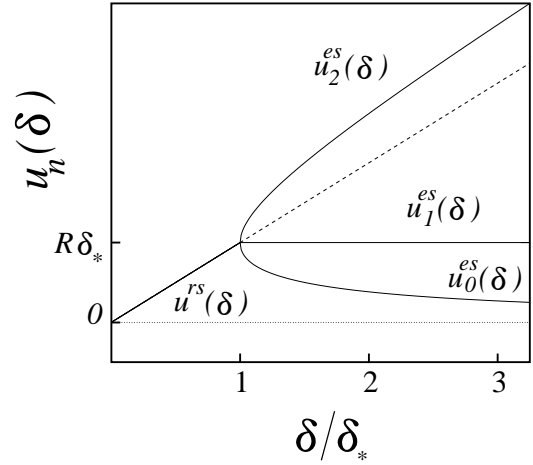}}
\end{picture}
\caption[]{\label{ui_paper} The saddle-point solutions  $ u_{n}$ of a system
with three degrees of freedom
as a function of the barrier  parameter $ \delta/\delta_{*}$.
For $ \delta < \delta_{*} $ the system escapes rigidly 
from the local minimum
of the potential via a configuration where all the particles are sitting on 
top of the barrier, $ u_{n} = u^{rs} $.
At $\delta= \delta_{*} $ the saddle splits and the 
elastic regime is entered for
$ \delta > \delta_{*}$. With increasing $ \delta $, $ u_{0}=u_{0}^{es}$ 
approaches the minimum, $ u_{1} =u_{1}^{es}= 
R \delta_{*}$ and the last particle 
$ u_{2} =u_{2}^{es} \to R(\delta + \delta_{*}) $ 
is hanging over the maximum of the single-particle potential.}
\end{figure}
Inserting the solutions for the elastic saddle $ {\bf q}_{es} $
into $ {\cal E} $, we can represent the free energy as a function of 
$ \epsilon $;
\begin{equation}
{\cal E}({\bf q}_{es}) = {\tilde U}_{B} (3-3\epsilon^{2}+\epsilon^{3}). 
\end{equation}
At $ \epsilon=0$ one finds
$ {\cal E}({\bf q}_{es}) = {\cal E}({\bf q}_{rs}) $.
For $ \epsilon > 0$, the value of $ {\cal E}({\bf q}_{es}) $  is smaller 
than that of $ {\cal E}({\bf q}_{rs})$. Thus there is a smooth crossover 
from the rigid $ {\bf q}_{rs}$to the elastic configuration 
$ {\bf q}_{es}$, which becomes the most probable one. 
To summarize, the activation energy of a
three particle chain is given by 
\begin{eqnarray}
U_{a}^{rs} & = & 
       3 U_{B} \delta^{3}, \\
U_{a}^{es} & = &        
         U_{B} \left( \delta^{3}
           + 3 \delta_{*}^{2} \delta
               - \delta_{*}^{3} \right) \nonumber
\end{eqnarray}
in the rigid and  elastic regimes, respectively. 
In order to visualize the most probable configuration leading to decay,
we represent the saddle-point solution in the original coordinates 
$ u_{0}, u_{1} $ and $ u_{2}$ as a function of the parameter  $ \delta$.
We find that for $ \delta > \delta_{*} $
\begin{eqnarray}
u_{0}^{es}  &=&  \frac{R}{2} \left[ \delta + \delta_{*}  
            - \left( \delta^{2} + 2 \delta  \delta_{*} 
               - 3  \delta_{*}^{2}  \right)^{1/2}  \right],\\
u_{1}^{es}  &=&  R \delta_{*}, \\
u_{2}^{es}  &=&  \frac{R}{2} \left[ \delta + \delta_{*}  
            + \left( \delta^{2} + 2 \delta  \delta_{*} 
               - 3  \delta_{*}^{2}  \right)^{1/2}  \right].
\end{eqnarray}
Note that there exists a second solution with the same energy, which can
be found by simply exchanging the indices $ 0$ and $2$.
\begin{figure}[t]
\unitlength1cm
\begin{picture}(8,4)
\epsfxsize=7.5cm
\put(0.5,0.5){\epsfbox{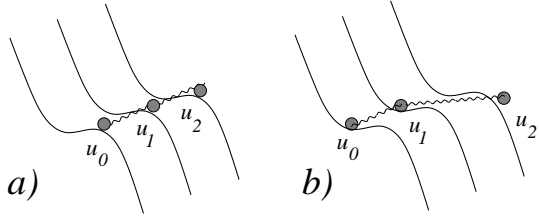}}
\end{picture}
\caption[]{\label{SPS_paper} (a) Rigid saddle-point solutions.
(b) Elastic saddle-point solutions.}
\end{figure}
The results are displayed in Fig.\ \ref{ui_paper} and illustrated in  
Fig.\ \ref{SPS_paper}.
By increasing the barrier parameter $ \delta $ above $ \delta_{*}$, 
the symmetry along the defect is broken as the elastic
saddle-point solution develops.  When   $ \delta $ is raised
further, particle $0 $ approaches the potential minimum at $ u_{min}=0$.
Particle $ 1 $ tries to adjust between its neighbors.
It is dragged  toward the minimum by particle $ 0 $, but, due to the  
coupling to particle $ 2$, there will be a finite distance between the 
particles
$ 1 $ and $ 0 $. On the other hand, particle $ 2 $ has swapped to the other 
side of the maximum.

\begin{figure}[t]
\unitlength1cm
\begin{picture}(8,5.5)
\epsfxsize=7.5cm
\put(0.5,0.5){\epsfbox{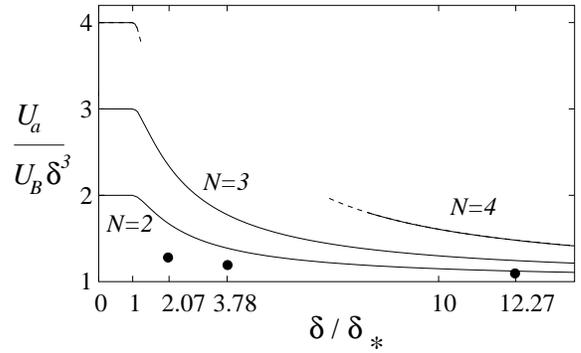}}
\end{picture}
\caption[]{\label{activation} 
The activation energy $ U_{a}$ normalized to 
the activation energy of a single particle $ U_{B} \delta^{3}$
 as a function of the barrier 
parameter $ \delta$ for various number of particles $N$.
For $ N=2,3$ the results are exact, for $ N>3$ the 
activation energy is calculated perturbatively in the
crossover regime $ \delta \sim \delta_{*}$ and in 
the limit of large $ \delta$.  
The activation energy for $ N=2$ and the experimental data (full dots) 
are taken from Ref.\ 5. } 
\end{figure}

Far in the elastic regime, $\delta / \delta_{*}\gg N^{2} $, 
we can generalize this picture to arbitrary $ N$.  
Making the ansatz $ u_{N-1} \gg u_{N-2} \gg u_{N-3} \sim 0$ 
we find the approximate solutions of $ \nabla {\cal E} = 0$, 
\begin{eqnarray}
&&u_{N-1}^{es}  \approx  R \delta + \kappa R^{3}/6U_{B},   \label{uNas1}  \\
&&u_{N-2}^{es}  \approx  \kappa R^{3}/6U_{B},\label{uNas2}\\
&&u_{n \le N-3}^{es} \approx 0\label{uNas3},
\end{eqnarray}
and the equivalent saddle $ u_{n} \to u_{N-1-n}$,
with an  activation energy
\begin{equation} \label{U_bound}
U_{a}  =  U_{B}  \delta^{3} \left(1+ \frac{\kappa R^{2} }{2U_{B} \delta}
 \right).
\end{equation}
The activation energy $ U_{a} $ is displayed in Fig.\ \ref{activation} 
for $ N=2,3,$ and $4$. 
Note that in this limit the elasticity term 
$ \kappa R^{2} \ll 2U_{B} \delta $ and the activation energy 
resembles that of a single particle $ U_{a} \sim U_{B} \delta^{3}$
with a renormalized barrier parameter. 
This means that for large $ \delta $ the system cannot gain much 
energy by nucleating at the boundary and bulk excitations 
become important. The bulk saddles are particle like 
excitations at position $m$ with a double kink,
\begin{eqnarray}
&&u_{m}^{bs}  \approx  R \delta + \kappa R^{3}/3U_{B}, \label{uNbs1}  \\
&&u_{m \pm 1}^{bs}  \approx  \kappa R^{3}/6U_{B},\label{uNbs2} \\
&&u_{n}^{bs}  \approx  0,\label{uNbs3} 
\end{eqnarray}
where $ |m-n| >1$ . They have an activation energy
\begin{equation} \label{U_bulk}
U_{a}   \approx U_{B}  \delta^{3} \left(
1 + \frac{\kappa R^{2} }{U_{B} \delta} \right),
\end{equation} 
which is larger than the activation energy of the elastic boundary
saddles.  Though energetically not preferable, for 
$ N \gg 1$ the decay can occur via bulk 
saddle-point solutions if the barrier parameter exceeds a crossover 
value $ \delta > \delta_{bs}$. The crossover to this new regime will 
be discussed in more detail in Sec. \ref{sec_pre}.

\section{Prefactor} \label{sec_pre}
Having determined the activation energies $ U_{a}(\delta) $ for the different 
regimes, the remaining task is to calculate the prefactor $ P(\delta) $
in Eq.\ (\ref{thermal_rate}).
Rewritten
in terms of $ {\bf q} = (q_{0},\dots,q_{N-1})$,  Eq.\ (\ref{thermal_rate1})
reads
\begin{equation} \label{thermal_rate2}
\Gamma_{th} = 
\sqrt{ \frac{U_{B}   k_{B}T |\tilde \mu_{0}^{s}|} 
{2 \pi N \eta^{2} R^{4} \delta} }
\frac{\int_{-\infty}^{\infty} d^{N-1} {\bf q^{'}}   
~{\rm e}^{  -{\cal E}( {\bf q^{'}})  /k_{B}T } }
{\int_{-\infty}^{0} d q_{0}   \int_{-\infty}^{\infty} d^{N-1} {\bf q^{''}}  
~{\rm e}^{  -{\cal E}( {\bf q})  /k_{B}T } }.
\end{equation}
Here $ {\bf q^{'}} = (q_{0}^{s},q_{1},\dots,q_{N-1}) $ is running along 
$ {\cal S}$  and $ {\bf q} = (q_{0},{\bf q^{''}})$ is probing $ {\cal V}$. 
In the denominator, $ q_{0} <0$  ensures that the integration is only 
performed over stable
configurations. The additional prefactor 
arises when transforming the integrals to the $q$ system
and taking into account that 
$ \mu_{0}^{s} = U_{B} \delta {\tilde \mu_{0}^{s}}/R^{2}$.

\subsection{Far from the crossover: Gaussian approximation}
In the Gaussian approximation, the integrals 
in the numerator and in the denominator in Eq. (\ref{thermal_rate2})
are evaluated by taking into account only the quadratic fluctuations 
around the saddle point  $ {\bf q}_{s}$, 
\begin{equation}
{\cal E}( {\bf q}) \approx   {\cal E}(  {\bf q}_{s} )
+  \frac{N {\tilde U}_{B}}{2}\sum_{k=0}^{N-1} {\tilde \mu_{k}}^{s} 
(q_{k}-q_{k}^{s})^{2},
\end{equation}
and the local minimum 
$ {\bf q}_{min}$,
\begin{equation}
{\cal E}( {\bf q} ) \approx  {\cal E}(  {\bf q}_{min} )
      +  \frac{N {\tilde U}_{B}}{2}\sum_{k=0}^{N-1} {\tilde \mu_{k}}^{min} 
(q_{k}-q_{k}^{min})^{2},
\end{equation}
respectively.
Thus one obtains a prefactor
\begin{eqnarray} \label{Gaussian_P}
P &=& \sum_{s} 
\frac{U_{B} \delta | {\tilde \mu_{0}}^{s} |  }{2 \pi  \eta R^{2}}
\left(  \prod_{n=0}^{N-1} \frac{  
                         {\tilde \mu_{n}}^{min} }
                 {     |{\tilde \mu_{n}}^{s}|} \right)^{1/2} \nonumber \\
&=& \sum_{s}\frac{| \mu_{0}^{s} |  }{2 \pi \eta}
 \left[ \frac{ \det H({\bf u}_{min}) }
                 {     |\det H({\bf u}_{s}) | } \right]^{1/2}, 
\end{eqnarray}
where the sum over the saddle index $s$ takes into account the 
contributions of equivalent saddles.
Here $  ({\tilde \mu_{n}^{min}}) \mu_{n}^{min} $ and 
$ ({\tilde \mu_{n}^{s}}) \mu_{n}^{s}  $ are 
the (dimensionless) eigenvalues of the curvature 
matrices $ H({\bf u}_{min})$ and $ H({\bf u}_{s})$ evaluated at the local 
minimum $({\bf q}_{min}) {\bf u}_{min}  $ and the saddles 
$ ({\bf q}_{s})  {\bf u}_{s}$, respectively. 
In contrast to a system with translational invariance, in the finite systems
considered here there is no Goldstone mode of the critical nucleus. 
Hence, well above and below the crossover, where $ \mu_{1}^{s} \not= 0 $, 
the evaluation of $ P $ is not corrupted by divergences.

In the rigid regime, we take  only the energetically lowest-lying saddle 
into account, and the sum over $s$ reduces to a single contribution.
With the determinants $\det H({\bf u}_{min}) $ and $ \det H({\bf u}_{rs}) $
given in Eqs.\ (\ref{Dmin_N}) and (\ref{Drs_N})  in Appendix \ref{appendix_a},
we find
\begin{equation} \label{rate_rigid}
 P(\delta < \delta_{*}) = \frac{ 3 U_{B} \delta  }{ \pi \eta R^{2}} 
\left[ \frac{\sinh ( N \Omega ) \tanh(\Omega/2)} 
            { \sin (N {\tilde \Omega} ) \tan( {\tilde \Omega}/2) } 
\right]^{1/2},
\end{equation} 
where $ \Omega  = 2~{\rm arcsinh}(\omega /2)$ 
and $ {\tilde \Omega}  = 2 \arcsin (\tilde \omega /2)$
with $ \omega = {\tilde \omega} =\sqrt{6 U_{B} \delta / \kappa R^{2}} $.
Below the crossover, two equivalent low-energy saddle-point solutions arise,
as was discussed in Sec.\ \ref{sec_sad}. The sum over both 
saddles gives rise to the factor $2$ in 
\begin{equation}  \label{rate_elastic}
P(\delta > \delta_{*})  = 2 \frac{| \mu_{0}^{es} |  }{2 \pi \eta}
 \left[ \frac{ \det H({\bf u}_{min}) }
                 {     |\det H({\bf u}_{es}) | } \right]^{1/2}.
\end{equation} 
In Eqs.\ (\ref{Del_N}) and  (\ref{muel})   we have estimated 
the determinant $ \det H({\bf u}_{es})$ and the eigenvalue 
$ \mu_{0}^{es}$, respectively, 
in the limit $ \delta \gg \delta_{*}$. We obtain
\begin{equation}  \label{pre_elastic}
P(\delta \gg \delta_{*}) \approx  
\frac{ 6 U_{B} \delta  }{ \pi \eta R^{2}}
\left[1 + {\cal O}(\delta_{*}/\delta) \right]. 
\end{equation}

As already mentioned in Sec.\ III, for $ N \gg 1 $ a crossover
to a regime can occur, where the decay dominantly occurs via bulk excitations.
The number of DOF $ N_{bs}$, where the crossover 
from boundary to bulk nucleation occurs, is found by comparing
the corresponding rates according to Eq.\ (\ref{thermal_rate}).
In the bulk regime, one has approximately $ N $ equivalent saddles and thus
with Eqs. (\ref{DNbs}) and (\ref{mubs}) the 
prefactor is given by
\begin{equation} \label{pre_bulk}
P  \approx  
N \frac{ 3 U_{B} \delta  }{ \pi \eta R^{2}}
\left[1 + {\cal O}(\delta_{*}/\delta) \right]. 
\end{equation} 
Comparing the rates for boundary and bulk nucleation 
with $ U_{a}$ given by  Eqs.\ (\ref{U_bound}) and (\ref{U_bulk}),
and $ P $ given by
Eqs.\ (\ref{pre_elastic}) and (\ref{pre_bulk}),
respectively, we obtain 
\begin{equation}
 \delta_{bs} \approx 
\left[ \frac{2 k_{B}T  \ln(N/2)}{\kappa R^{2}} \right]^{1/2}.
\end{equation} 
Note, that within our approximations
the choice of the system specific  parameters 
$ N,R,\kappa $  and the temperature $ T$ is restricted
to values that meet the constraint $ \delta_{bs} \ll 1$.

\subsection{Near the saddle-point bifurcation: Beyond steepest descent}
\begin{figure}[t]
\unitlength1cm
\vspace{1cm}
\begin{picture}(8,5)
\epsfxsize=7.5cm
\put(0.0,0.5){\epsfbox{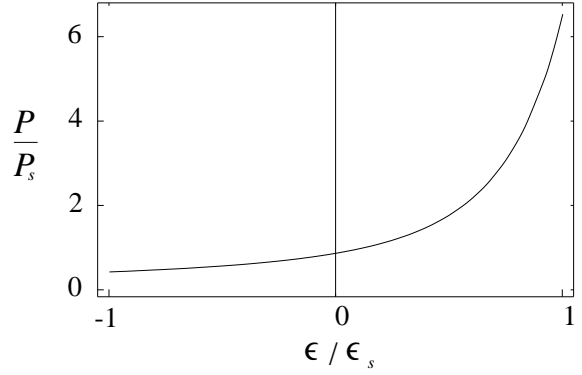}}
\end{picture}
\caption[]{\label{scaling} Scaling property of the prefactor $ P $
as a function of $ \delta $ near $\delta_{*}$. 
$ P/P_{s} $ is shown as a 
function of the distance $ \epsilon$ from the crossover.
Both $ P_{s} $ and $ \epsilon_s$ are system specific scaling variables.} 
\end{figure}
In the crossover regime, where $ \delta \to \delta_{*}$ and hence 
$ |\epsilon| \to 0$, the prefactor calculated in the Gaussian approximation 
diverges as $ P \sim 1/\sqrt{\epsilon}$
due to  the vanishing eigenvalue $ {\tilde \mu_{1}} =-3 \epsilon$.
The divergence can be regularized 
by taking into account the third order terms in $ q_{1}$
in the approximation of $ {\cal E}({\bf q}')$ around the saddle point
in Eq.\ (\ref{thermal_rate2}).
Defining the system-dependent scaling variables 
\begin{displaymath}    
P_{s}  \approx 
 \frac{\left[54 \tanh ( \Omega /2 )~  \sinh(N \Omega) \right]^{1/2}
U_{B}^{7/4} \delta_{*}^{9/4} 
  }
{\pi^{3/2} \eta  R^{3}  (N k_{B}TD)^{1/4}  \tan(\pi/2N)}
\end{displaymath} 
and
\begin{displaymath}
\epsilon_{s} = \left( \frac{16 k_{B}T D}{9 N \delta_{*}^{3} U_{B} } 
\right)^{1/2},
\end{displaymath}
we show in Appendix \ref{appendix_pre} that
\begin{equation}   \label{crossfun}
P(\epsilon) = P_{s} F(\epsilon/\epsilon_{s}),
\end{equation} 
where the function $ F $ is found to be
\begin{equation}
F(y) = \left\{
         \begin{array}{ll}
        \pi  \sqrt{ \frac{|y|}{2}}
   \exp(y^{2}) 
   \left[ I_{-1/4}(y^{2}) 
   - I_{1/4}(y^{2}) \right],
 & \delta < \delta_{*}, \\
\\
         8^{-1/4}\Gamma(1/4), & \delta = \delta_{*}, \\
\\
        \pi  \sqrt{ \frac{|y|}{2}}
     \exp(y^{2}) 
   \left[ I_{-\frac{1}{4}}(y^{2}) 
   + I_{\frac{1}{4}}(y^{2}) \right],
 & \delta > \delta_{*}.
         \end{array}
       \right. 
\end{equation} 
For large $ |\epsilon/\epsilon_{s}|$ the prefactor given in 
Eq.\ (\ref{crossfun}) matches with the Gaussian result. 
However, in the crossover regime, where $ |\epsilon/\epsilon_{s}| < 1 $,
the Gaussian prefactor deviates strongly from Eq.\ (\ref{crossfun}),
as expected, since here the Gaussian approximation becomes invalid.
Since we considered a metastable situation, where 
$ k_{B}T \ll U_{B} \delta^{3} $, we have $ \epsilon_{s} \ll 1 $.
Hence, the crossover regime is extremely narrow, 
$ |\delta-\delta_{*}| \ll \delta_{*}$.
The function $ F=P/P_{s}$, which is shown in Fig.\ \ref{scaling},
reflects  two interesting aspects.
First, one realizes that the behavior of the rate is smooth at the 
crossover. The divergences that occur in the Gaussian approximation 
are regularized by taking into account higher orders of the 
fluctuation coordinates. Second, $F$  can be regarded as a scaling function,
where the constants $ \epsilon_{s}$ and $P_{s}$
contain the system-specific parameters. 
The scaling relation is universal in the sense that it does not depend on the
details of the considered system. Of course, a constraint is that the 
crossover must be of second order to guarantee the validity of the 
perturbative treatment that we applied. However, we have excluded
systems with a single-particle potential that enforce a first-order transition
from the beginning. 
Note, that Eq.\ (\ref{crossfun}) was found by
taking into account only the cubic terms of the 
modes $ q_{0},~ q_{1}$, and $q_{2}$.
These long-wavelength excitations determine the 
decay process at the crossover, where the discreteness of the system becomes 
irrelevant. Hence the
result can be applied to continuous systems as well. In fact, a similar
crossover function is found at the second-order transition from thermal 
to quantum decay of a single particle in a metastable state \cite{Grab87}. 
Formally, this theory can also be used to describe a 
rigid-to-elastic crossover in the thermal decay of an 
elastic line escaping from  a homogeneous defect, 
but with {\em periodic} instead of {\em open}
boundary conditions, which we considered here. Note that the scaling function 
found in Ref.\ 19 is different from ours. One can indeed show,
that the functional form of the scaling function is influenced by 
the symmetry of the system.

\section{Discussions and Conclusions}
We studied the thermal decay of a chain of 
elastically coupled particles from a metastable state.
The metastability arises from each of the particles being trapped in
a local minimum of their single-particle potential. The energy barrier
that separates the local minimum from energetically lower-lying ones 
can be tuned by a barrier parameter $ \delta $. At $ \delta = 0 $ the
energy barrier vanishes and the metastability ceases to exist.
With increasing $ \delta$, we find three regimes. For small $ \delta $, 
the decay occurs mainly via a rigid configuration, where all the DOF 
leave the trap at once. 
At $ \delta_{*} =  2 \kappa R^{2} \sin^{2} (\pi/2N) / 3 U_{B}\ $
a saddle-point bifurcation occurs, which marks a
crossover from  rigid to  elastic motion.
For $\delta > \delta_{*} $ the decay occurs mainly via boundary nucleation.
However, at even higher values $ 1 \gg \delta > \delta_{bs} > \delta_{*} $
a crossover to bulk nucleation can take place.

Our main goal was to evaluate the thermal decay  rate 
$ \Gamma_{th} = P \exp(-U_{a}/k_{B}T)$
in the three regimes. This involves the calculation of the prefactor $ P$
and the activation energy $ U_{a}$. 
The latter is given by the energy  $ {\cal E}$ of the most probable configuration 
leading to decay, namely, the lowest-lying saddle-point solution.
We solved the problem  for $ N=3$ particles exactly.
Furthermore, we treated the case of an arbitrary 
number $N$ of DOF  perturbatively in the crossover regime and deep in the
elastic regime. We have shown
how the system uses its elasticity to lower the activation energy in the 
elastic regime. Whereas in the rigid regime 
the activation barrier is $ U_{a}^{rs} = N U_{B} \delta^{3}$, in the
elastic regime near the crossover 
$ U_{a}^{es}\approx U_{a}^{rs}(1-C \epsilon^{2})$, where 
$ \epsilon =  1 - \delta_{*}/ \delta$ and $ C \sim 1$ 
is a positive constant
that depends on the details of the potential. 
Increasing  $\delta$ in the elastic regime, the 
particles first escape via nucleation at the boundaries with an activation 
energy $ U_{a}^{es} \sim U_{B} \delta^{3} +\kappa R^{2} \delta^{2}/2$,
where the first term arises from the potential energy of the activated 
particle and the second term is the elastic energy of the kink that occurs in 
the boundary saddle. Due to the imposed free (von Neumann) boundary 
conditions, this kind of activation is energetically preferred 
compared to bulk nucleation with an activation energy
$ U_{a}^{bs} \sim U_{B} \delta^{3} +\kappa R^{2} \delta^{2}$.
Since the bulk saddle consists of two kinks, twice the
elastic energy is needed to activate a bulk nucleation process.
However, in large systems, with $ N \gg 1$, bulk nucleation becomes more probable
for $ 1\gg \delta > \delta_{bs} = \sqrt{ 2 k_{B}T \ln(N/2)/ \kappa R^{2} } $.
Above $ \delta_{bs}$ the many possibilities to excite a 
particle somewhere in the bulk, which grow as $ N$
in the prefactor $P$, outnumber the two 
possibilities of boundary nucleation.
At large $ \delta$, the elastic interaction between the particles
becomes less and less important and the activation energy approaches 
the energy $ U_{B} \delta^{3}$ which is needed to excite a single particle  
over the barrier independently of the others.
To discuss the relevant energy scales, we now  
fix all variables except $N$. The crossover occurs when the number of 
DOF is increased above $ N_{bs} = 2 \exp(\kappa R^{2} \delta^{2}/2k_{B}T)$.
Hence, when the elastic coupling is weak and the temperature is high,
bulk nucleation already occurs at lower values of $N_{bs}$.  
The crossover is thus determined by the ratio of elastic energy and 
thermal energy.

Second, we determined the prefactor $P$.
Far from the rigid-to-elastic crossover, 
the calculation of the prefactor $ P$ was done in
Gaussian approximation both in the rigid and elastic regimes. 
Near the crossover, the Gaussian approximation breaks down due to
a diverging integral, which is caused by a vanishing eigenvalue of the 
curvature matrix. 
By taking into account higher orders in the fluctuation coordinates, we
remove the divergence and obtain a smooth behavior of the rate at the 
crossover. The prefactor of the rate exhibits a scaling 
property $ P/P_{s} = F(\epsilon/\epsilon_{s})$.
The function $F$ is universal, but depends on the symmetries 
of the model. The scaling parameters $ P_{s}$ 
and $ \epsilon_{s}$ are system-specific constants.

At the saddle-point bifurcation  
$ U_{a}(\delta),$ $ U_{a}'(\delta),$ $ P(\delta),$ $  P'(\delta), $ and 
$  P''(\delta) $ are continuous, whereas $U_{a}''(\delta) $ is discontinuous.
Hence $ \Gamma_{th}(\delta)$ and $ \Gamma_{th}'(\delta)$ are continuous,
but $ \Gamma_{th}''(\delta)$ is discontinuous. Interpreting $ U_{a}$ as a
thermodynamic potential, one easily sees the analogy between the crossover
described here and a second-order phase transition. This analogy becomes even 
clearer when the integral in the
enumerator in Eq.\ (\ref{thermal_rate1}) is interpreted as the reduced 
partition sum over the DOF transverse to the unstable direction.  
Note that close to the crossover the discrete structure of the model
becomes unimportant, this kind of crossover can also be also found in 
continuous systems \cite{Chri95,Sima90,Cast96}.
The question arises whether first-order-like transitions could occur 
also in the thermal decay of elastic chain systems. 
As in the crossover from thermal to quantum decay \cite{Lark83,Chud92}
the type of the crossover depends crucially on the shape of the 
single-particle potential $ U(u_{n})$. For a cubic parabola as is 
discussed in this work, the crossover is of second order. However, one
could imagine other physical systems where the single-particle potential
has a form that causes a first order transition.

The discrete model that we have used here is quite general.
In the following we will discuss the application of the theory to 
two physical situations, the dynamics of the phases in DJTL's
 and the thermal creep of 
pancake vortices in layered superconductors with columnar defects.

DJTL are parallelly coupled one
dimensional Josephson-junction arrays, 
and the $N$ DOF in this case are 
the phase differences across each of the $N$ Josephson junctions.  
In current driven DJTL, metastable states occur when the DOF are trapped
in a local minimum of the tilted washboard potential common to these systems. 
For $N = 2$, the problem reduces to the decay of the phases in a
current biased dc SQUID \cite{Ivle87,Lefr92,Mora94}. Both the rigid decay
\cite{Han89}, where the two phases behave as a single one, and the
elastic case  \cite{Lefr92}, where the two phases decay one after 
another, were experimentally observed. 
In the continuous limit, $N \to \infty $, the system becomes identical to
a long JJ. The rigid-to-elastic 
crossover occurs \cite{Sima90,Cast96} when the junction length $ L_{J} $ 
becomes of the order of the 
Josephson length $ L_{J} \sim \pi \lambda_{J}$.
Here we analyzed a model for a DJTL, that provides 
a system to study the intermediate case of decay from a metastable 
state with  a finite number of DOF. An experimental investigation of the 
rigid-to-elastic 
crossover requires that the current $I$ can be driven through the crossover 
current $I_{*} = N I_{c} (1-\delta_{*}^{2}) $. 
An orientation for the choice of the system parameters can be obtained by 
comparison with the dc-SQUID \cite{Lefr92,Han89}, noting that 
$I_{*} - N I_{c} \propto h^{2}c^{2}/ ( e^{2 }L^{2} I_{c}^{2} N^{4})$. 
A systematic experimental study of the rigid-to-elastic crossover as a 
function of the system parameters $ L,~ I_{c}, $ 
and $ N $ is still lacking and would be highly desirable.
A remaining question was, if additional 
crossovers occur in systems with a large number of DOF.
In addition to the rigid-to-elastic crossover 
due to a saddle-point bifurcation
of the potential energy, we find that in systems with large $N$
a second crossover from boundary to bulk nucleation 
can take place.   DJTL's with a large number
of DOF offer the possibility to observe such a crossover 
by varying system-specific parameters or the temperature.

Let us now discuss our theory in the context of a single stack of pancake 
vortices  trapped in a columnar defect in a layered superconductor.
In the presence of a current density $j$ that flows within the 
layers, the vortices are  driven by the resulting  Lorentz force.
Once thermally activated from the defect, the pancake stack starts to
move through the sample until it is trapped by another defect. The resulting
motion is called thermal vortex creep.
A typical example for a layered system is a high-temperature 
superconductor (HTSC). A HTSC like  YBCO  is characterized by
an anisotropy  $ \gamma \sim 5$, and the ratio of the penetration
depth to the coherence length is $ \lambda_{ab}/\xi_{ab} \sim 100$. 
The distance between the layers and their thickness are  
$s \sim t \sim \xi_{ab}$, and the defect radius is $ R \sim 2 \xi_{ab}$. 
In order to observe the transition 
from rigid to elastic decay experimentally, the ratio 
$ (j_{c} -j_{*})/j_{c}>0$ must be sufficiently large. 
However, substituting the defect energy 
 $ U_{B} \sim t \varepsilon_{0} \ln(R/\xi_{ab})$ and 
the elastic energy  $ \varepsilon_{l} R^{2}/s $, with 
$ \varepsilon_{l} 
= (\varepsilon_{0}/\gamma^{2}) \ln(\lambda_{ab} / \xi_{ab})$, 
into $(j_{c} -j_{*})/j_{c} = \delta_{*}^{2}$,
one finds that 
even in systems with low anisotropy and a small number of layers  
 $ (j_{c} -j_{*})/j_{c} < 10^{-2}$, 
indicating that the phenomenon could hardly be
observed  experimentally in high-$T_c$ superconductors since
$ j_{*}$ is very close to $ j_{c}$.
Thus, for large currents $j_{c} -j \ll j_{c} $ as considered here,
the vortex system turns out to be mainly in the elastic regime where  
the layered structure of the material is important.
Then, the activation barrier $ U_{a} $ is of
the order of the single-particle barrier $ U_{B}(1-j/j_{c})^{3/2}$, 
which  can be interpreted as a 
vortex creep induced by the escape of individual
pancakes from the columnar defect \cite{Bran92a,Nels92}. 
This ``decoupling'' regime can be also entered from the low-current half-loop 
regime $j \ll j_{c}$, when the width of the bulk critical nucleus 
becomes of the order of the layer separation \cite{Kosh96}.
We find that at low temperatures $T$ the thermal creep is induced
by boundary (surface) nucleation. It would be interesting to investigate
experimentally if the crossover from bulk to surface  nucleation might be 
observed in thin layered samples.  
In sum, we calculated analytically the creep rate for coupled particles 
trapped in a metastable state and found that an interesting behavior 
arises from the interplay between elasticity, pinning, discreteness and 
finite-size effects. 

\section*{Acknowledgments}
We indebted to   H.~Schmidt, J.~K\"otzler,  G.~Blatter, 
O.~S.~Wagner, A.~V.~Ustinov and A.~Wallraff 
for fruitful discussions. Financial support from the DFG-Projekt
No.~Mo815/1-1 and the 
Graduiertenkolleg ``Physik nanostrukturierter Festk\"orper,'' University of
Hamburg, is gratefully acknowledged.

\begin{appendix}
\section{Determinant and eigenvalues of the curvature matrix} 
\label{appendix_a}

\subsection{Recurrence relation for the Hessian matrix}
As was shown in Sec.\ \ref{sec_pre}, the prefactor $ P $ of the thermal decay
rate is a function of the determinant and the eigenvalues of the 
curvature matrix evaluated at the relative minimum and the saddle points,
respectively, see Eq.\ (\ref{Gaussian_P}).
The curvature or Hessian matrix ${\bf H}_{N}  $ with matrix elements 
$ 
H_{nm}({\bf u}_{0}) = \partial_{n}\partial_{m} {\cal E}({\bf u}_{0})
$
determines the nature of $ {\cal E} $ at the extremum 
$ {\bf u}_{0} $. If all eigenvalues of  ${\bf H}_{N}({\bf u}_{0})  $     
are negative (positive), $ {\bf u}_{0} $ is a relative maximum (minimum).
If some of the eigenvalues are positive and some are negative, then 
$ {\bf u}_{0} $ is a saddle point.
For ${\cal E}({\bf u}_{0})$ with $N \ge 3$, the Hessian matrix reads
\begin{displaymath}
{\bf H}_{N}({\bf u}) =\left(
\begin{array}{cccccc}
 \partial_{0}^{2} {\cal E}({\bf u})& -\kappa & 0      & \cdots  & 0   & 
-\alpha \kappa\\
-\kappa & \ddots& \ddots &         &         & 0      \\
    0   & \ddots & \ddots & \ddots  &         & \vdots \\
 \vdots &        & \ddots & \ddots  & \ddots  & \vdots \\
    0   &        &        & \ddots  & \ddots  & -\kappa \\
-\alpha \kappa &    0   & \cdots &  0      & -\kappa & \partial_{N-1}^{2} 
{\cal E}({\bf u})
\end{array}
\right).
\end{displaymath}
In the case of open boundary conditions
$ \alpha=0$, the diagonal elements are given by 
\begin{displaymath}
\partial_{n}^{2} {\cal E}({\bf u})
= 
\left\{
   \begin{array}{cc}
     \kappa + U'' (u_{n}), & n=0,N-1         \\
     2 \kappa   + U'' (u_{n}),       & 0 < n < N-1. 
   \end{array}
\right.
\end{displaymath}

In the discussion that follows, we introduce  
\begin{equation} \label{D_N}
 D_{N} =\det_{N}\left(
     \begin{array}{cccccc}
        1+ x_{0} & -1     &   0    & \cdots         & 0       \\
          -1   & 2+ x_{1} &   -1   &  \cdots         & \vdots  \\
        \vdots & \ddots        & \ddots & \ddots         & \vdots  \\
        \vdots & \ddots        & \ddots & 2 + x_{N-2} & -1      \\
           0   & \cdots        &   0    &      -1        & 1+ x_{N-1}
     \end{array}
    \right),
\end{equation}
which is used to calculate both the determinant and the
characteristic polynomial of ${\bf H}_{N}$.
For example, in order to calculate the determinant of the  
normalized Hessian ${\bf H}_{N}/\kappa$ for $N>4$, 
one sets $ x_n =   U''(u_{n}) / \kappa$.
Below, we will derive a recurrence relation, which is used to determine
$ D_{N}$ in some special cases.

By shifting the last column to the first and then lifting the bottom row to 
the top, one can rewrite the determinant as 
\begin{displaymath}
 D_{N} =\det_{N}\left(
     \begin{array}{ccccccc}
    1+ x_{N-1} & 0     &   0    & \cdots &   0    &  -1 \\
    0   &    1+x_{0}  &   -1   &  0    & \cdots &   0 \\
    0   &    -1    &  &  & &   \\
 \vdots &      0   &  &  & &  \\
   0  &   \vdots &  &  A_{N-2}& &  \\ 
   -1    &   0      &  &  & &   
      \end{array}
    \right),
\end{displaymath}
where the $ (N-2) \times (N-2)$ matrix $A_{N-2}$ is given by
\begin{displaymath} 
 A_{N-2} = \left(
     \begin{array}{cccccc}
        2+ x_{1} & -1     &   0    & \cdots         & 0       \\
          -1   & 2+ x_{2} &   -1   &  \dots         & \vdots  \\
           0   & \ddots        & \ddots & \ddots         &        0   \\
        \vdots & \ddots        & -1     & 2+ x_{N-3} & -1      \\
           0   & \cdots        &   0    &      -1        & 2+x_{N-2}
     \end{array}
    \right).
\end{displaymath}
In the following, we will consider the case where 
$  x_{1} = \dots = x_{N-2} =x$. Note that $ x_{0}$ and $ x_{N-1}$ can be 
arbitrary.

Expanding  $ D_{N}$, we find with $G_{n}=\det A_{n}$
\begin{eqnarray} \label{DN1}
D_{N} & = &
(1+x_{N-1}) \left[
                 (1+x_{0}) G_{N-2} - G_{N-3}
\right] \nonumber \\
&&- (1+x_{0}) G_{N-3} +  G_{N-4} .
\end{eqnarray}
Expanding the determinant $G_{n}  $ according to the last row of $ A_{n}$, 
one finds the recursive relation \cite{Gelf60}
$ G_{n} = (2+x_{n}) G_{n-1} - G_{n-2}$
that can be rewritten as a difference equation
\begin{equation} \label{difference}
( G_{n} - G_{n-1} ) - ( G_{n-1} - G_{n-2} )  
-x_{n} G_{n-1} = 0.
\end{equation}
The initial conditions are given by the determinants $ G_{1} $ 
and $ G_{2} $,
\begin{eqnarray} 
G_{1} & = &  2+x, \nonumber \\ \label{init}
G_{2} & = & (2+ x)^{2}-1.
\end{eqnarray}
For $ 2 \le N \le 4$,
we can use the recurrence relations for $ G_{n}$,
if we define $ G_{0} =1$,  $ G_{-1}  =0$,
and  $ G_{-2} =-1$.

\subsection{Uniform case}
The solution of these difference equations
is possible for special cases. We now analyze the uniform
case where  $ x= x_{0} = \dots = x_{N-1}$.
Then  Eq.\ (\ref{DN1}) simplifies to
\begin{equation} \label{DNhomo}
D_{N}  =  (1+x)^{2} G_{N-2} - 2(1+x) G_{N-3} + G_{N-4}. 
\end{equation}

\subsubsection{Determinant at the relative minimum, $ x \ge 0$}
We first discuss the case of the local minimum $ {\bf u} ={\bf u}_{min} $,
where $x= \omega^{2} >0 $.
Imposing the initial conditions given by Eq.\ (\ref{init}), one obtains a
solution \cite{Gelf60} of Eq.\ (\ref{difference}),
\begin{equation} \label{Gmin} 
G^{min}_{N-1} = \frac{\sinh(N \Omega)}{\sinh \Omega},
\end{equation}
where 
\begin{displaymath} 
\sinh \frac{\Omega}{2} = \frac{\omega}{2}. 
\end{displaymath}
Using Eqs.\ (\ref{DNhomo}) and  (\ref{Gmin}), we obtain
\begin{equation} \label{Dmin_N}
D^{min}_{N} = \omega^{2}  G_{N-1}^{min} = 
2 \tanh \left(\frac{ \Omega }{2} \right)~  \sinh (N \Omega) .
\end{equation}

\subsubsection{Determinant at the rigid saddle, $x<0$}
In the same way as for the local minimum, one obtains $ D_{N} $
at the rigid saddle $ {\bf u} ={\bf u}_{rs} $ but now with negative 
$x = - {\tilde \omega}^{2}<0$. One finds
\begin{displaymath}
G^{rs}_{N-1} = \frac{\sin(N {\tilde \Omega})}{\sin {\tilde \Omega}},
\end{displaymath}
where
\begin{equation} \label{omega2}
\sin \frac{{\tilde \Omega}}{2} = \frac{{\tilde \omega}}{2}, 
\end{equation}
and hence
\begin{equation} \label{Drs_N}
D^{rs}_{N} = 2 \tan\left(\frac{ {\tilde \Omega} }{2}\right)~   
\sin (N {\tilde \Omega})  
\end{equation}

\subsubsection{Eigenvalues}
The eigenvalues of ${\bf H}_{N}$ are found by evaluating the roots of the
characteristic polynomial, $ \det( {\bf H}_{N} - \mu {\bf I} )=0$.
We have again a determinant of the form of Eq.\ (\ref{D_N}), but now
with $ x_{n} =   U''(u_{n})/\kappa  -\mu /\kappa $, such that we can define 
$ D_{N}(\mu) =  \kappa^{-N} \det( {\bf H}_{N} - \mu {\bf I})$.
Using Eq.\ (\ref{Drs_N}) we find that the
roots where $ D_{N}(\mu) = 0 $ 
are given by $ {\tilde \Omega}_{m} = m \pi/N$, 
where $ m=0,\dots,N-1 $. 
Inserting $ {\tilde \Omega}_{m} $ into Eq.\ (\ref{omega2}) 
yields ${\tilde \omega}_{m} = 2 \sin({\tilde \Omega}_{m}/2)$,
hence $ D_{N}(\mu_{m}) = 0 $ for
\begin{equation} \label{eigenss}
   \mu_{m} = 
     4 \kappa\sin^{2} \left( \frac{m \pi}{2N} \right) 
     + U''(u_{0}),     
\end{equation} 
which are the eigenvalues of 
$  {\bf H}_{N}({\bf u}_{0}) $  for a given uniform extremal solution
${\bf u}_{0}=(u_{0},\dots,u_{0})$.

\subsection{Nonuniform case}
Approximate solutions for the determinant and the eigenvalues 
can be obtained deep in the elastic
regime, $ \delta/ \delta_{*} \gg 1$. 

\subsubsection{Elastic boundary saddle 
$(\delta_{bs} > \delta \gg \delta_{*}) $}
For the elastic boundary saddle-point 
configurations obtained in Eqs.\ (\ref{uNas1})-(\ref{uNas3}),
to highest order in $ \delta/ \delta_{*} $ one finds
that $ U''(u_{0}) = \cdots =  U''(u_{N-3}) \approx 6 U_{B} \delta /R^{2} $,
$  U''(u_{N-2}) \approx 6 U_{B} \delta /R^{2} -2 \kappa $, and
$ U''(u_{N-1}) \approx - 6 U_{B} \delta /R^{2} -2 \kappa $.

With $ x_{n} = U''(u_{n})/\kappa $ one obtains for the determinant 
up to $ {\cal O}\left( \delta^{N-2} \right) $
\begin{eqnarray} \label{Del_N}
D_{N}^{es} & \approx & (1+x_{N-1})(2+x_{N-2})(1+x_{0})  G_{N-3}^{min}.
\end{eqnarray}
The ratio $ D_{N}^{min}/D_{N}^{es}$, 
which is needed to calculate the prefactor in the elastic regime
is found to be
\begin{displaymath} 
  \frac{D_{N}^{min}}{D_{N}^{es}} = -1 - \frac{\kappa R^{2}}{3 U_{B} \delta}
 + {\cal O}\left[ (\delta_{*}/\delta)^{2} \right].
\end{displaymath}

To calculate the eigenvalues, we
set again $ x_{n} =  U''(u_{n})/\kappa - \mu/\kappa$. 
The characteristic polynomial 
$ D_{N}(\mu) $ 
is now up to ${\cal O}( \delta^{N-2})$, given by
\begin{displaymath}
D_{N}(\mu) \approx   (1+x_{N-1})(2+x_{N-2})(1+x_{0}) G_{N-3}(\mu).
\end{displaymath}
Thus, to lowest order in $ \delta $, we find that the
smallest eigenvalue is
\begin{eqnarray} \label{muel}
\mu_{0}^{es} & \approx & = -\kappa  - \frac{6 U_{B} \delta}{R^{2}} .
\end{eqnarray}

\subsubsection{Elastic bulk saddle $(\delta > \delta_{bs}) $}
For the elastic bulk saddle-point 
configurations obtained in Eqs.\ (\ref{uNbs1}) and (\ref{uNbs2})
to highest order in $ \delta/ \delta_{*} $ one finds 
for a double kink situated at $ m$,
$ U''(u_{m}) \approx - 6 U_{B} \delta /R^{2} -4 \kappa $,
$  U''(u_{m \pm 1}) \approx 6 U_{B} \delta /R^{2} -4 \kappa $, and
for $ |n-m| > 1$ $ U''(u_{n})  \approx 6 U_{B} \delta /R^{2}$.
With $ x_{n} = U''(u_{n})/\kappa $ and using periodic boundary 
conditions, the determinant is approximately given by
\begin{displaymath} 
D_{N}^{bs}  \approx   (2+x_{m+1})(2+x_{m})(2+x_{m-1}) G_{N-3}^{min}.
\end{displaymath}
The ratio $ D_{N}^{min}/D_{N}^{bs}$ is 
\begin{equation} \label{DNbs}
  \frac{D_{N}^{min}}{D_{N}^{bs}} = -1 - \frac{4\kappa R^{2}}{3 U_{B} \delta}
 + {\cal O}\left[ (\delta_{*}/\delta)^{2} \right].
\end{equation}
The characteristic polynomial 
$ D_{N}(\mu) $ 
is now up to ${\cal O}( \delta^{N-2})$ given by
\begin{displaymath} 
D_{N}(\mu) \approx   (2+x_{m+1})(2+x_{m})(2+x_{m-1}) G_{N-3}(\mu),
\end{displaymath}
where $ x_{n} =  U''(u_{n})/\kappa - \mu/\kappa$. 
Thus, to lowest order in $ \delta $, we find that the
smallest eigenvalue is
\begin{eqnarray} \label{mubs}
\mu_{0}^{bs} & \approx & = -2\kappa  -\frac{6 U_{B} \delta}{R^{2}} .
\end{eqnarray}

\section{Prefactor in the crossover regime} \label{appendix_pre}
\subsection{Rigid regime
$(\delta {\mathrel{\raise.4ex\hbox{$<$}\kern-0.8em\lower.7ex\hbox{$\sim$}}} 
\delta_{*})$}
For $ \delta \to \delta_{*}$, both the eigenvalue $ \mu_{1}^{rs} $ and
the determinant $ D_{N}^{rs} $ vanish.
Hence the Gaussian integral containing $ \mu_{1}^{rs}$ in 
Eq.\ (\ref{Gaussian_P})
diverges, and third-order terms in $ q_{1} $ have to be taken into account.
In the rigid regime, the third-order expansion of $ {\cal E}$ in $q_{1}$
is given by Eq.\ (\ref{beyond1}).
The contributions to $ P $ of all degrees of freedom except 
$ q_{1} \in {\cal S}$ are found by Gaussian integration:
\begin{eqnarray} 
P & = & \frac{U_{B} \delta    }{ 2 \pi  \eta R^{2}}
\left(   \frac{ | {\tilde \mu_{0}}^{rs} | \prod_{n=0}^{N-1}
                         {\tilde \mu_{n}}^{min} }
                 {  \prod_{n = 2}^{N-1}    {\tilde \mu_{n}}^{rs}} \right)^{1/2}
   \left(\frac{N U_{B} \delta^{3}}{2 \pi k_{B}T} \right)^{1/2} \nonumber  \\
&& \times \int_{-\infty}^{\infty} dq_{1}
\exp \left[ -\frac{N {\tilde U}_{B}}{2k_{B}T}  
\left( {\tilde \mu_{1}^{rs}} q_{1}^{2}
 + D q_{1}^{4}  \right)   \right].
\end{eqnarray} 
In the following we first derive an approximate expression
for 
$ \prod {\tilde \mu_{n}}^{min}  / 
\prod_{n \not= 1} {\tilde \mu_{n}}^{rs}  $ and then evaluate the 
remaining integral over $ q_{1}$.

For the calculation of the product term we use the relation
$ \prod {\tilde \mu_{n}}^{min}  / 
\prod_{n \not= 1} {\tilde \mu_{n}}^{rs} = 
{ \tilde \mu_{1}}^{rs} D_{N}^{min} / D_{N}^{rs} $. 
Let us analyze $ D_{N}^{rs} $ for $ {\tilde\mu}_{1}^{rs}$ close to zero.
Recall that
\begin{displaymath}
{\tilde \omega}^{2} = -\frac{U''(u_{0})}{\kappa} 
= 4 \sin^{2} \left( \frac{\pi}{2N} \right) 
     - \frac{\mu_{1}^{rs}}{\kappa}.
\end{displaymath}
Inserting this expression into Eq.\ (\ref{omega2}) 
in the limit of small $\mu_{1}^{rs} $,
we find
\begin{displaymath}
{\tilde \Omega} \approx \frac{\pi}{N} - \frac{\mu_{1}^{rs}}
                                             {2 \kappa \sin (\pi/N)},
\end{displaymath}
such that, to lowest order in $ \mu_{1}^{rs}$,
\begin{displaymath}
\sin (N {\tilde \Omega}) \approx  \frac{N \mu_{1}^{rs}}
                                             {2 \kappa \sin (\pi/N)},
\end{displaymath}
and
\begin{displaymath}
\tan \left(\frac{\tilde \Omega}{2}\right) 
\approx  \tan  \left(\frac{\pi}{2N}\right).
\end{displaymath}
Hence
\begin{displaymath}
 { \tilde \mu_{1}}^{rs}\frac{ D_{N}^{min}} {D_{N}^{rs}}
= - \frac{4 \kappa}{N} \cos^{2} \left( \frac{\pi}{2N} \right) 
   \tanh \left(\frac{\Omega}{2} \right) \sinh (N \Omega ) .
\end{displaymath}

The integration over  $ q_{1}$ yields
\begin{eqnarray} 
\int_{-\infty}^{\infty} dq_{1}
\exp \left[ -\frac{N {\tilde U}_{B}}{2k_{B}T}  
\left( {\tilde \mu_{1}^{rs}} q_{1}^{2}
 + D q_{1}^{4}  \right)   \right] \nonumber \\ 
= \frac{1}{2} 
    \sqrt{
       \frac{ {\tilde \mu_{1}^{rs}} }
          { D }}
    \exp  \left[
     \frac{ N {\tilde U}_{B}({\tilde \mu_{1}^{rs}})^{2}} 
          {16k_{B}T D }
          \right] 
    K_{1/4} \left[
     \frac{N {\tilde U}_{B}({\tilde \mu_{1}^{rs}})^{2}} 
          {16k_{B}T D }
          \right],
\end{eqnarray}
where $ D $ as defined above in Eq.\ (\ref{D}) 
arises during the Gaussian integrations over 
$ q_{0}$ and $q_{2}$. $ K_{1/4} $ is the modified Bessel function.
We make the substitution $ {\tilde \mu_{1}^{rs}}=-3 \epsilon$. 
After defining
\begin{eqnarray} \label{A}
P_{s} &=& \left(  \frac{ U_{B}^{2} \delta^{2}  | {\tilde \mu_{0}^{rs}} | 
\prod_{n=0}^{N-1} {\tilde \mu_{n}}^{min} }
{ 8 \pi^{3}\eta^{2} R^{4} 
\prod_{n=2}^{N-1}  {\tilde \mu_{n}^{rs}}} \right)^{1/2}
\left(\frac{N \delta^{3} U_{B}}{ k_{B}TD} \right)^{1/4} \nonumber \\
& \approx & 
 \frac{\left[54 \tanh ( \Omega /2 )~  \sinh(N \Omega) \right]^{1/2}
U_{B}^{7/4} \delta_{*}^{9/4} 
  }
{\pi^{3/2} \eta  R^{3}  (N k_{B}TD)^{1/4}  \tan(\pi/2N)} 
\end{eqnarray}
and 
\begin{equation} \label{B}
\epsilon_{s} = \left( \frac{16 k_{B}T D}{9 N \delta_{*}^{3} U_{B} } 
\right)^{1/2},
\end{equation}
which are constants to leading order in $ \epsilon $, we obtain
the prefactor of the rate for the rigid region of the  crossover 
regime 
$ \delta {\mathrel{\raise.4ex\hbox{$<$}\kern-0.8em\lower.7ex\hbox{$\sim$}}} 
\delta_{*}:$,
\begin{equation} \label{P_rigid}
P(\epsilon)= 
\frac{\pi P_{s}}{\sqrt{2}}~~ \sqrt{\left| \frac{ \epsilon}{\epsilon_{s} } 
\right|} 
\left[ I_{-1/4}\left(\frac{ \epsilon^{2}}{\epsilon_{s}^{2}} \right) 
- I_{1/4}\left(\frac{ \epsilon^{2}}{\epsilon_{s}^{2}} \right)  \right]
\exp\left(\frac{ \epsilon^{2}}{\epsilon_{s}^{2}} \right).
\end{equation}

\subsection{Elastic regime 
$(\delta {\mathrel{\raise.4ex\hbox{$>$}\kern-0.8em\lower.7ex\hbox{$\sim$}}} 
\delta_{*})$} In the elastic regime near the crossover, where 
$ \epsilon 
{\mathrel{\raise.4ex\hbox{$>$}\kern-0.8em\lower.7ex\hbox{$\sim$}}} 0$, 
we expand $  {\cal E}({\bf q})$  around the
perturbative elastic saddle-point solution (\ref{approx_ela}), 
\begin{displaymath}
{\cal E}({\bf q}) =  {\cal E}({\bf q}_{es}) 
+ \frac{1}{2} {\cal E}^{(2)}(\{\xi_{k}\})
 + \frac{1}{6} {\cal E}^{(3)}(\{\xi_{k}\}),
\end{displaymath}
where $ {\cal E}^{(2)}$ and ${\cal E}^{(3)}$ contain the 
terms of second and third order, respectively, and
$ \xi_{k} = q_{k} -q_{k}^{es}  $
are the fluctuations around the elastic saddle point.
By introducing the shifted fluctuation coordinates for $m \not = 1$,
\begin{displaymath}
{\hat \xi_{m}} = 
\xi_{m} + \frac{2 q_{1}^{es} \xi_{1} A_{m} }{{\tilde \mu_{m}}^{rs}},
\end{displaymath}
with $ A_{0} =-3,~ A_{2}=-3/2$ and $ A_{i>2}=0$, we find,
for the quadratic part to leading order in $\epsilon$,
\begin{displaymath}
{\cal E}^{(2)}
= - 2 {\tilde \mu_{1}^{rs}} \xi_{1}^{2} 
  + \sum_{m \not = 1} {\tilde \mu_{m}^{rs}} {\hat \xi_{m}}^{~2}. 
\end{displaymath}

Note that $ {\tilde \mu_{m}^{rs}}$ are the dimensionless
eigenvalues evaluated at the {\em rigid} saddle-point configuration.
Within the crossover regime, to leading order in $ \epsilon$, the eigenvalues 
at the elastic saddle-point solution
$ {\tilde \mu_{i \not= 1}^{es}} = {\tilde \mu_{i \not= 1}^{rs}} $ 
are independent of $ \epsilon$, except 
$ {\tilde \mu_{1}^{es}} =-2{\tilde \mu_{1}^{rs}}=2 \epsilon /3 $.
The higher order contributions to the expansion read
\begin{displaymath} 
\frac{1}{6}  {\cal E}^{(3)}  =  
\left( \sum_{m \not = 1} A_{m} {\hat \xi_{m}} \right)   \xi_{1}^{2}
+2D q_{1}^{es} \xi_{1}^{3}.
\end{displaymath} 
Transforming the fluctuation coordinates a second time,
\begin{eqnarray}
{\tilde \xi_{m \not = 1} } 
& = & {\hat \xi_{m}}  + \frac{A_{m}}{{\tilde \mu_{m}}^{rs}} \xi_{1}^{2}, \nonumber \\
{\tilde \xi_{1}} &  = &  \xi_{1} + q_{1}^{es}, \nonumber
\end{eqnarray} 
we find
\begin{displaymath}
{\cal E}({\bf q}) = {\cal E}({\bf q_{es}})  +
\frac{1}{2}  \sum_{m \not = 1}{\tilde  \mu_{m}}^{rs} {\tilde \xi_{m}}^{~2}  + 
\frac{D}{2}  \left[ {\tilde \xi_{1}}^{~2} - 
\left(q_{1}^{es} \right)^{2} \right]^{2}.
\end{displaymath}
By using $ (q_{1}^{es})^{2}=-\mu_{1}^{rs}/2D = 3 \epsilon/2D$,
we evaluate the integrals as in the previous paragraph, \cite{Grad:3.548}  
\begin{eqnarray}
&& \int_{-\infty}^{\infty} d{\tilde \xi_{1}} 
\exp \left\{ -\frac{D}{2k_{B}T} 
\left[ {\tilde \xi_{1}}^{~2} - (q_{1}^{es})^{2} \right]^{2}  \right\}
  \nonumber  \\
&& = \frac{\pi}{2 \sqrt{2} } 
            ~\sqrt{ \left| \frac{ {\tilde \mu_{1}}^{rs} } { D } \right| }
 ~\left\{
  I_{-1/4} \left[ \frac{({\tilde \mu_{1}}^{rs})^{2}}{16k_{B}T D } 
\right]
 +
  I_{1/4} \left[ \frac{({\tilde \mu_{1}}^{rs})^{2}} {16k_{B}T D } 
\right] 
 \right\} \\ 
&& \times
    ~\exp  \left[  -\frac{({\tilde \mu_{1}}^{rs})^{2}}{16k_{B}T D } \right], 
\nonumber
\end{eqnarray} 
where $ I_{1/4}$ and  $ I_{-1/4}$ are modified Bessel functions.
The prefactor of the rate for the elastic regime 
$ \delta 
{\mathrel{\raise.4ex\hbox{$>$}\kern-0.8em\lower.7ex\hbox{$\sim$}}} 
\delta_{*} $ in the crossover region then reads
\begin{equation} \label{P_elastic}
P(\epsilon)= 
\frac{\pi P_{s}}{\sqrt{2}}~~ \sqrt{ \frac{ \epsilon}{\epsilon_{s} }} 
\left[ I_{-1/4}\left(\frac{ \epsilon^{2}}{\epsilon_{s}^{2}} \right) 
+ I_{1/4}\left(\frac{ \epsilon^{2}}{\epsilon_{s}^{2}} \right)  \right]
\exp\left(\frac{ \epsilon^{2}}{\epsilon_{s}^{2}} \right).
\end{equation}

\end{appendix}

\end{document}